# Double metal-insulator transitions and magnetoresistance: An intrinsic feature of Ru substituted $La_{0.67}Ca_{0.33}MnO_3$

L. Seetha Lakshmi,[*,@] V. Sridharan,[*] R. Rawat,[§] A. A. Sukumar,[‡] M. Kamruddin,[*]
V. S. Sastry[*] and V. S. Raju[‡]

[*]Materials Science Division, Indira Gandhi Centre for Atomic Research, Kalpakkam, Tamil Nadu 603 102, India
[§] UGC-DAE Consortium for Scientific Research, Khandwa Road, Indore, Madhya Pradesh 452 017, India
[‡]National Centre for Compositional Characterization of Materials, Hyderabad, Andra Pradesh -500 062, India

**ABSTRACT**

In this paper, we examine the possible influence of extrinsic factors on the electrical and magnetotransport of $La_{0.67}Ca_{0.33}Mn_{1-x}Ru_xO_3$ (x ≤ 0.10). Ru substitution results in two metal-insulator transitions, both exhibiting magnetoresistance (MR): a high temperature sharp maximum at $T_{MI1}$ followed by a relatively broad maximum at lower temperatures (LTM) at $T_{MI2}$. The two transitions shift to lower temperatures at a rate of ~ 3 K/at.% and ~ 16 K/at.% respectively. No additional magnetic signal corresponding to $T_{MI2}$ could be observed either in ac susceptibility or in specific heat. Preparation history is same for all the compounds and the final sintering is carried out 1500°C for 36 hrs in flowing oxygen. Grain size of the compounds is in the range ~ 18,000-20,000 nm which is not small enough to warrant a LTM in the electrical transport. Absence of any additional peaks in the high statistics powder XRD not conforming to orthorhombic P*nma* symmetry, linear systematic increase in the lattice parameters and the unit cell volume, close matching of the transition temperatures in resistivity, ac susceptibility and specific heat and their linear systematic decrease with increasing x and a homogeneous distribution of Mn, Ru and O at arbitrarily selected regions within the grain and across the grains exclude another possibility of chemical inhomogeneity as being the cause for the LTM. The insensitiveness of grain boundary magnetoresistance to Ru composition indicates the grain boundary is not altered upon Ru substitution to result in a LTM. Oxygen stoichiometry of all the compounds is close to the nominal value of 3. These results not only exclude the extrinsic factors, but also establishes that double metal transitions both exhibiting MR is *intrinsic* to Ru substituted $La_{0.67}Ca_{0.33}MnO_3$ and the system. These results substantiate our hypothesis that Ru substituted system undergoes a *magnetic phase separation involving the co-existence of two ferromagnetic-metallic phases in its ground state*.

## I. INTRODUCTION

The last decade has seen a renaissance of research activities focused towards the complex ground state properties of the ortho-perovskite manganites.[1,2,3,4,5] Among the most important aspects unveiled is the presence of intrinsically inhomogeneous states in nano or mesoscopic length scales, sensitive to the external magnetic field.[6] Theories based on the microscopic phase separation[7] emerged soon to provide a realistic starting point to the physics of manganites. The multitude of experimental data[8,9,10] poured out in the recent past indeed have established the mixed-phase tendencies in the manganites, consistent with the theoretical predictions. Though many factors are still unexplored or currently under discussion, it is believed that the phase separation in manganites could result from atleast two mechanisms: (i) electronic phase separation, where the competing states have different hole densities and 1/r Coulomb effects lead to co-existence of phases at length scales of nano meters (ii) disorder–driven phase separation near first-order transitions, with competing states of equal density that leads to large (sub-μm) co-existing clusters. These models propose that the competitive magnetic interactions in the background of disorder effects lead to colossal magnetoresistive effects in manganites.[11]

Since the essential degrees of freedom; lattice, spin, charge and orbital; are closely associated with the Mn ion, Mn site substitution studies offer a direct method of manipulating the disorder effects. Moreover, suitable paramagnetic substitutions offer an additional handle to probe the local spin coupling, an unexplored parameter of the rare-earth substitution studies. There are numerous works addressing the importance of Mn site substitutions with paramagnetic ions such as $Ni^{2+}$,[12] $Cr^{3+}$,[13,14] $Ru^{4+/5+}$,[15,16] $Rh^{3+}$[17] in inducing the colossal magnetoresistance (CMR) in the charge ordered (CO) insulating manganites. In these systems, just a few percent substitution (~ 3 at.%) at the Mn site is reported to be sufficient for a collapse of the CO state and appearance of the ferromagnetic state, while ~ 7-27 T field is required to melt the CO antiferromagnetic state and to induce the CMR effect.[18,19] Mn site substitution of $Pr_{0.5}Ca_{0.5}MnO_3$ is found to generate two different classes of materials[20] - spin glass like insulators upon diamagnetic and paramagnetic substitutions result in metallic ferromagnets exhibiting CMR phenomena. Among the paramagnetic substitutions such as $Ni^{2+}$,[12] $Co^{3+}$,[21] $Cr^{3+}$,[13,22,23] $Ir^{3+}$,[24] $Rh^{4+}$,[17] $Ru^{4+/5+}$,[15,16] Ru ion is reported to be most efficient in melting the CO state and inducing the ferromagnetism and metallicity in zero magnetic field.[25] For instance, the metal-insulator transition temperature of Ru substituted compound is reported to be as high as 240 K, while for others, it was found to be typically less than 150 K.[20] Hence, Ru appears to be

---

[@] Corresponding Author: Email: slaxmi73@gmail.com (L. Seetha Lakshmi)

exceptional among the paramagnetic substitutions which may provide an insight into the nature of local spin coupling effect on the transport and magnetic properties of CMR manganites.

In our previous studies on the Ru substituted $La_{0.67}Ca_{0.33}MnO_3$ compound, it was found that Ru substitution results in the least reduction in the transition temperatures[26] compared to other paramagnetic substituents.[27,28,29,30,31] From the lowest reduction in the transition temperatures, it was speculated that the local spin coupling between Ru and neighboring Mn ion is ferromagnetic and such a local ferromagnetic coupling partially compensates the deleterious local structural effects.[26] Interestingly, Ru substitution results in two metal-insulator transitions[32] (MITs) both exhibiting peaks in the magnetoresistance.[33] Based on the structural, magnetic and electrical and magnetotransport studies of Ru substituted system and from the inter-comparison of the earlier works, it was speculated that the double maxima in the magnetotransport is closely associated to the MITs. However, as will be discussed in Section III the double maxima is also shown to arise from the extrinsic factors such as synthesis conditions, chemical inhomogeneity, grain size, grain boundary effects and oxygen off-stoichiometry. To check these possibilities, we have carried out micro structural studies: SEM and elemental mapping using EDAX, magnetoresistance as a function of applied field at 5 K and thermogravimetric studies and the results will be presented in this paper. These results, as will be discussed subsequently, exclude the extrinsic factors as being the cause of the second low temperature maximum in the electrical and magneto-transport. The double maxima in the magnetotransport establish the fact that it is intrinsic to Ru substituted $La_{0.67}Ca_{0.33}MnO_3$. This strongly substantiates our hypothesis that the Ru substituted $La_{0.67}Ca_{0.33}MnO_3$ undergoes a magnetic phase separation involving the co-existence of two ferromagnetic-metallic (FM-M) phases in its ground state. Though further microscopic experimental studies are in progress, we argue, based on the current knowledge accumulated on the manganite perovskites that such a magnetic phase separation, distinctly different from the electronic phase separation is intrinsic to paramagnetic substituted systems with ferromagnetic-metallic ground state. These results are expected to provide a starting point for further theoretical investigation in this area.

## II. EXPERIMENT

Polycrystalline samples of $La_{0.67}Ca_{0.33}Mn_{1-x}Ru_xO_3$ (x ≤ 0.10) were prepared by solid state reaction. The final sintering was carried out in single batch at 1500 °C for 36 hrs in flowing oxygen. Sufficient care was taken at different stages of the sample preparation to ensure that the preparation history is identical for all samples prepared. The room temperature powder XRD patterns with a step size of 0.05° in the 2θ range 15-120° were recorded in Bragg-Brentano para-focussing geometry with $Cu_{K\alpha}$ radiation using STOE diffractometer. Dwell time of 30 sec was chosen to achieve adequately high statistics (~$10^5$ counts for 100% peak at 2θ~32.7°). The XRD patterns were analyzed with the Rietveld method using the GSAS program.[34] Temperature variation of resistivity in zero field (ρ(T)) was measured in van der Pauw geometry.[35] ac susceptibility measurements were performed on powder samples using a home built ac susceptometer under an ac probe field of 0.25 Oe and an excitation frequency of 941 Hz. The in-phase component of ac susceptibility (χ′) alone is considered for further discussion. The magnetoresistance (MR) in an applied field of 5 T was measured using a standard four probe technique. A superconducting magnet was employed with the magnetic field direction parallel to the current direction. The field dependent MR measurements were also carried out at 5 K. Specific heat ($C_p$) measurements were carried out in the temperature interval 140-300K using a Differential Scanning Calorimeter (Mettler DSC821$^e$) calibrated for temperature and caloric scales with Ar purge gas. All compounds used had masses of approximately 40 mg. Sapphire was used as the reference material for $C_p$ evaluation. The sample microstructures were recorded on a Scanning Electron Microscope (Philips XL 30) equipped with a field emission gun at 25 keV. Elemental mapping of Mn, Ru and O were carried out at randomly selected areas within the grains and across the grains using an Energy Dispersive X-ray Analysis attachment with an ultra thin sapphire window Si (Li) detector with 135 eV resolution at 5.9 keV. The oxygen stoichiometry was estimated by decomposing the compounds at 1350°C, under a reducing atmosphere of Ar–4% $H_2$ mixture flushed at a controlled flow rate. Approximately 30 mg of the powder sample loaded in the recrystallized alumina crucible was heated from the room temperature at a linear programmed rate up to 1350°C in the reducing atmosphere and kept isothermally at 1350°C for 40 minutes. The thermogravimetric analyser (SETARAM, Model: SETSYS 16/18) coupled to a mass spectrometer was employed for these studies. The oxygen content of the starting compound was estimated from the weight loss (ΔW) at the end of the isothermal step assuming the following reduction reaction,

For undoped compound:

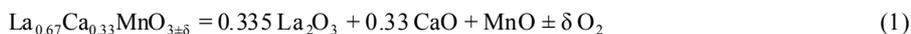
$$La_{0.67}Ca_{0.33}MnO_{3\pm\delta} = 0.335\,La_2O_3 + 0.33\,CaO + MnO \pm \delta\,O_2 \qquad (1)$$

For Ru substituted compounds:

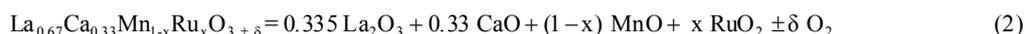
$$La_{0.67}Ca_{0.33}Mn_{1-x}Ru_xO_{3\pm\delta} = 0.335\,La_2O_3 + 0.33\,CaO + (1-x)\,MnO + x\,RuO_2 \pm \delta\,O_2 \qquad (2)$$

There is a close agreement between the percentage weight losses expected for the formation of the final products due to decomposition and those observed in the thermogram. Oxygen off-stoichiometry (δ) was calculated from ΔW according to the relation $\delta = \dfrac{\Delta W \times W_f}{W_i \times W_o}$ where $W_i$, $W_f$ and $W_O$ denote the initial weight of the compound, formula weight of the compound and molecular weight of oxygen respectively.

## III. RESULTS AND DISCUSSION

High statistics room temperature powder x-ray diffraction patterns of the compounds are shown in **FIG.1**. All the peaks could be indexed to orthorhombic structure with Pnma space group (No.62). The Rietveld refinement yielded an excellent agreement between the calculated and the observed patterns indicating that samples are single phase in nature. As a representative of the series, Rietveld refinement spectra of $x = 0.10$ is shown in the inset of **FIG.1**. It is seen that the lattice parameters a, b and c and thus cell volume (v) increases linearly with Ru in the entire range of substitution (Error! Reference source not found.). For instance, while a and b show a similar increase of ~ 0.43 % for $x = 0.10$, an increase of ~ 0.20 % is observed for c. The valence state of Ru in these compounds has been a subject of debate.[36,37,38] Among the possible valence states, the states pertinent to the present study are 3$^+$, 4$^+$ and 5$^+$ with an ionic radius of 0.68 Å, 0.62 Å and 0.565 Å respectively for a six-fold co-ordination.[39] There are reports, which propose the presence of Ru$^{4+}$ and Ru$^{5+}$ (iso-electronic to the DE pair, Mn$^{3+}$ and Mn$^{4+}$) ions in these compounds.[25,37] However, the presence of Ru$^{4+/5+}$ is expected to introduce a decrease in the lattice parameters. On the other hand, the appreciable increase observed in the present case (**FIG. 2**) indicates the other two possibilities. One such possibility is the presence of either Ru$^{4+}$ or Ru$^{3+}$ with Ru$^{4+}$ replacing Mn$^{4+}$ and Ru$^{3+}$ replacing Mn$^{3+}$ ion. However, inferences from magnetization,[32] electrical transport and magnetoresistance measurements, as discussed subsequently, do not favor the presence of Ru$^{3+}$ or Ru$^{4+}$ alone. This strengthens the other possibility, viz., Ru being in a mixed valence state of Ru$^{4+}$ iso-electronic to Mn$^{3+}$ and Ru$^{3+}$ being iso-electronic to Fe$^{3+}$ ion and such a combination could have lead to an increase in the unit cell parameters. X-ray photoemission spectroscopic study on Ru substituted (La-Sr)-Mn-O system of similar composition [38] has established the presence of $Ru^{3+}$ and $Ru^{4+}$ pair and this supports the above inference.

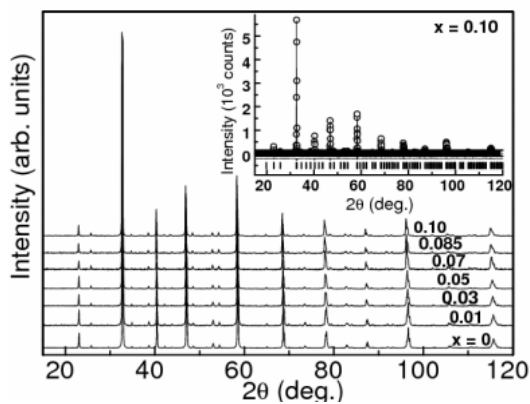

**FIG.1**: High statistics room temperature powder x-ray diffraction patterns of $La_{0.67}Ca_{0.33}Mn_{1-x}Ru_xO_3$ ($0 \leq x \leq 0.10$) compounds. Miller indices of the major Bragg-reflections are also indicated. Inset shows the Rietveld refinement pattern for $x = 0.10$, the highest Ru concentration of the present study. Symbol denotes the observed intensity, continuous lines denotes the calculated intensity. Difference between the observed and the calculated intensities are shown at the bottom of the figure. Positions for the calculated Bragg-reflected positions are marked by the vertical bars.

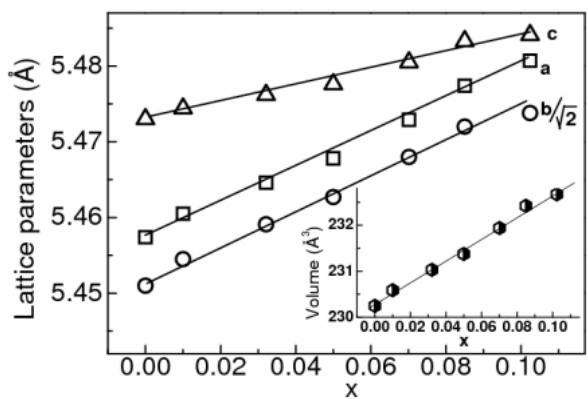

**FIG. 2**: Compositional dependence of lattice parameters ($a$, $b$ and $c$) (in Å) and unit cell volume (v) (in Å$^3$) (inset) of $La_{0.67}Ca_{0.33}Mn_{1-x}Ru_xO_3$ ($0 \leq x \leq 0.10$) compounds. Error bars are smaller than symbols. The line is guide to eye.

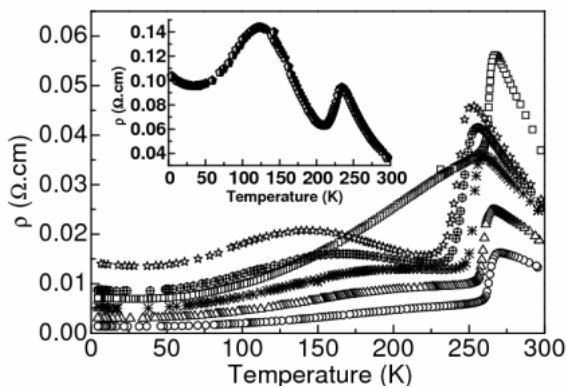

**FIG. 3**: Temperature variation of resistivity ($\rho(T)$) in the absence of magnetic field of $La_{0.67}Ca_{0.33}Mn_{1-x}Ru_xO_3$ compounds for $x = 0$ (□), 0.01 (○), 0.03 (△), 0.05 (✷), 0.07 (⊕), 0.085 (☆) For clarity, similar curve for 0.10 (☽) is given in the inset of the figure.

As the temperature is decreased from room temperature, the virgin compound, $La_{0.67}Ca_{0.33}MnO_3$, undergoes a metal to insulator transition (MIT) with a sharp maximum at $T = T_{MI}$ in $\rho(T)$ marking a transition from semiconductor like behavior

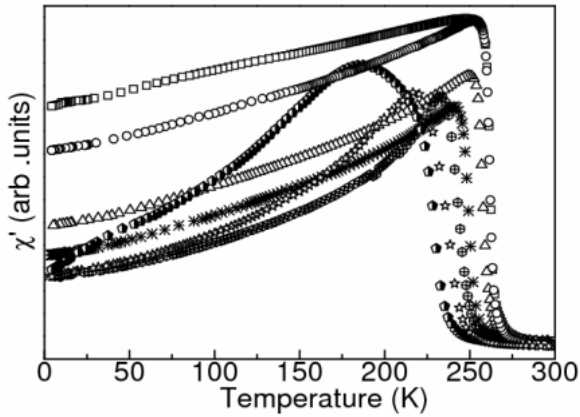

**FIG. 4 :** Temperature variation of in-phase component of ac susceptibility ($\chi'(T)$) of $La_{0.67}Ca_{0.33}Mn_{1-x}Ru_xO_3$ compounds (for x = 0 (□), 0.01 (○), 0.03 (△), 0.05 (✳), 0.07 (⊕), 0.085 (☆) and 0.10 (⬤)) under an ac probe field of 0.25 Oe and an excitation frequency of 941 Hz

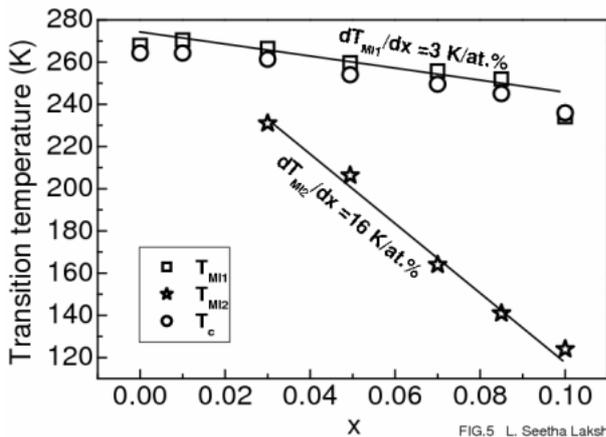

**FIG. 5:** Composition dependence of metal to insulator transition temperatures ($T_{MI1}$ and $T_{MI2}$) and para to ferromagnetic transition temperature ($T_c$) of $La_{0.67}Ca_{0.33}Mn_{1-x}Ru_xO_3$ ($0 \leq x \leq 0.10$) compounds. Straight lines are the best linear fit to the experimental data to estimate the rate of suppression in the transition temperatures.

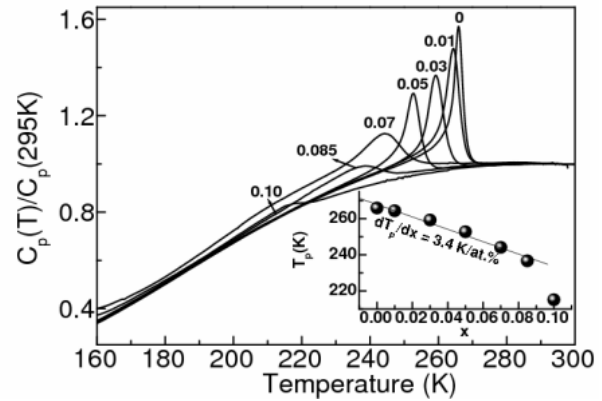

**FIG. 6:** Normalized $C_p$ ($(C_p(T)/C_p(295\,K))$) curves of $La_{0.67}Ca_{0.33}Mn_{1-x}Ru_xO_3$ ($0 \leq x \leq 0.10$) compounds. Inset shows the variation of peak temperature as a function of Ru concentration. Straight line is the best linear fit to the experimental data to estimate the rate of suppression in the transition temperature.

($\partial\rho/\partial T < 0$) to metallic behavior ($\partial\rho/\partial T > 0$) (**FIG.3**). Close to the MIT, all the compounds exhibit a para to ferromagnetic transition (PM-FM) at $T_c$ (**FIG. 4**). Interestingly, Ru substituted compounds exhibit two maxima in $\rho(T)$: the high temperature maximum at $T = T_{MI1}$ followed by a relatively broad maximum at still lower temperatures ($T = T_{MI2}$) (**FIG. 3**). It is worth mentioning that the second low temperature broad maximum (LTM) is perceptible for $x \geq 0.05$. Both the maxima shift systematically to lower temperatures with increasing Ru concentration. While the high temperature maximum at $T_{MI1}$ shifts at a rate $(dT_{MI1}/dx)$ of ~ 3K/at.% up to x = 0.085, the smallest rate reported for the Mn site substituted CMR manganites, and much more rapidly beyond that composition, the latter maximum shifts at a rate $(dT_{MI2}/dx)$ ~ 16 K/at.% (**FIG.5**), comparable to that of Fe substituted $La_{0.67}Ca_{0.33}MnO_3$ system.[26] This is in contrast to the charge ordered (CO) Ru substituted $La_{0.4}Ca_{0.6}MnO_3$,[40] where the LTM is reported not to shift with Ru concentration, but remained centred at 125 K. PM-FM associated to $T_{MI1}$ is also found to shift to lower temperatures with $dT_c/dx = 3.2$ K/at.%, no magnetic (ac susceptibility) signal corresponding to low temperature maximum could be observed (Error! Reference source not found.). The normalized specific heat curves of $La_{0.67}Ca_{0.33}Mn_{1-x}Ru_xO_3$ ($x \leq 0.10$) compounds are given in **FIG.6** The virgin compound gives rise to a sharp specific heat peak close to its magnetic transition ($T_c$). The specific heat anomaly shifts to lower temperature (**inset of FIG.6**) and becomes progressively broader with an increase of Ru concentration. These observations are in agreement with that of ac susceptibility measurement. No such anomaly

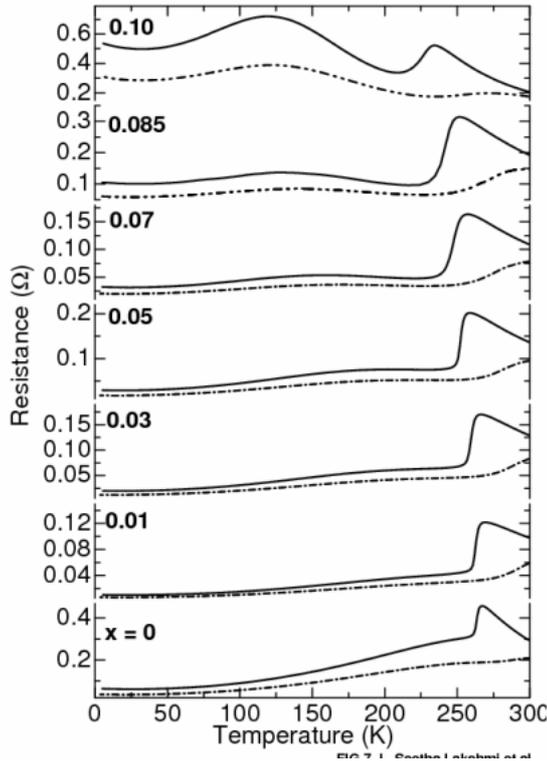

FIG. 7 : Temperature dependence of resistance in the absence (continuous line) and presence of magnetic field of 5 T (dotted line) of $La_{0.67}Ca_{0.33}Mn_{1-x}Ru_xO_3$ ($0 \leq x \leq 0.10$) compounds.

corresponding to $T_{MI2}$ could be observed in our measurements.

With the application of magnetic field of 5 T, an overall reduction in the resistance was observed (**FIG. 7**). As the magnetic field further broadens the inherently broad $T_{MI2}$, the maximum could not be determined for x = 0.01 and 0.03.

Though both maxima shift to higher temperature on application of magnetic field, a characteristic feature of CMR manganites, their field dependence is somewhat different. While, the high temperature maximum ($T_{MI1}$) shifts beyond the maximum temperature range of measurement for $x \leq 0.07$, the shift of $T_{MI2}$ progressively decreases and a negligible shift is observed for x = 0.10. The magnetoresistance (MR) in percentage is defined as

$$MR(\%) = \frac{(\rho_H - \rho_0)}{\rho_0} \times 100 \qquad (3)$$

where $\rho_H$ and $\rho_0$ are the resistivity of the compounds in the presence and absence of magnetic field. The temperature dependence of MR in a field H of 5 T ($MR_5(T)$) is shown in **FIG.8**. While the MR curve of the undoped compound exhibits single peak, MR curves of Ru substituted systems exhibit two peaks: a dominant peak close to its $T_{MI1}$ (in the zero field resistance) and a weak one close to its $T_{MI2}$ (in zero field resistance). Though the latter is arguably weak, it becomes prominent and broader with Ru concentration (**FIG.8**). A significant magnetoresistance with weak temperature dependence could also be seen well away from the MR peaks and this is shown to be a characteristic feature of the polycrystalline manganites.[41]

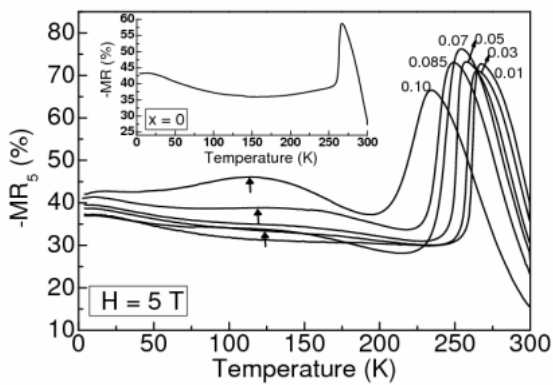

FIG. 8 : Temperature dependence of magnetoresistance at 5 T field ($MR_5(T)$) of $La_{0.67}Ca_{0.33}Mn_{1-x}Ru_xO_3$ ($0 \leq x \leq 0.10$) compounds.

Indeed, there are numerous reports on the presence of double maxima in the $\rho(T)$ curve and most of these reports infer the origin of the double maxima feature to extrinsic factors such as the synthesis conditions,[42] chemical inhomogeneity,[43,44] grain size,[45,46] grain boundary (GB)[47,48] and oxygen off-stoichiometry ($\delta$).[49] For the present work, it is pertinent to explore if the aforementioned factors are the cause of the LTM of Ru substituted $La_{0.67}Ca_{0.33}MnO_3$. Two maxima in the electrical transport have been observed in ceramics of differently doped La-mangnaites prepared by sol-gel,[45,50] solid state reactions,[51,52] combustion with urea[53] and wet chemical method.[54] Of these reports, only the result of Sun et. al[51] exhibits a sharp and distinct maximum in $\rho(T)$ just near $T_c$ as observed in Ru substituted compounds. Extreme care was taken at different stages of the synthesis of the compounds for the present investigation. The precursor materials $La_2O_3$ (Indian–Rare Earths), $CaCO_3$ (CERAC), $MnO_2$ (CERAC) and $RuO_2$ (CERAC) are of purity better than 99.99 %. $La_2O_3$ was pre-heated at 800°C for 24 hrs to remove the moisture before weighing. The compounds were given several heat treatments in the temperature range 1200-1400°C in flowing oxygen atmosphere with several intermediate grindings followed by pelletization. Sufficient care was taken at different stages so that the preparation history was same for all samples prepared. The pellets were reground and compacted using PVA as a binder to improve the inter-grain connectivity. PVA added pellets were ramped slowly in the temperature range, 200-600 °C and soaked for 3-6 hrs to remove the traces of PVA. The final sintering at 1500 °C for 36 hrs in flowing oxygen was carried out in single batch to ensure that all the samples were given the identical sintering conditions. Typical density was determined to be in the range 92-97 % of theoretical value and it shows no systematic variation with x. It is believed that the sample synthesis route and the sintering conditions are not the cause for the observed LTM in the Ru substituted compounds of the present study.

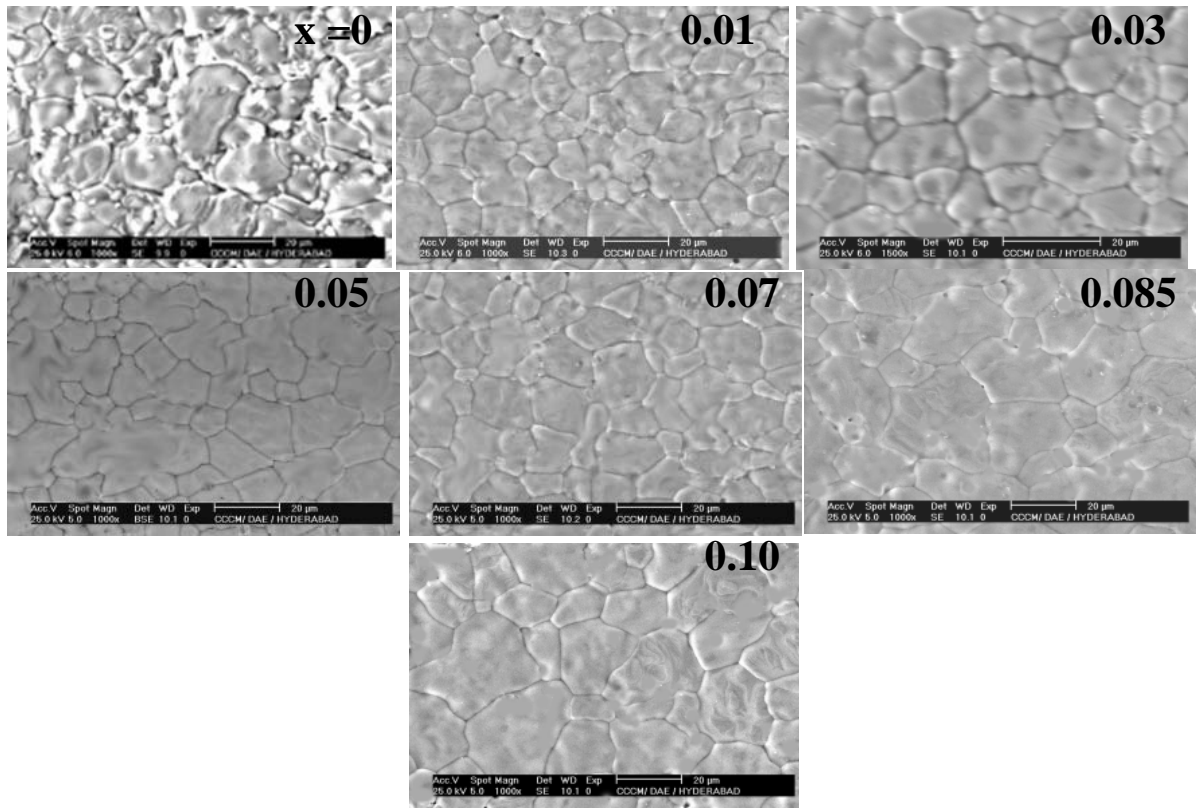

**FIG. 9**: SEM micrograph of $La_{0.67}Ca_{0.33}Mn_{1-x}Ru_xO_3$ ($0 \leq x \leq 0.10$) compounds.

Indeed quite a number of works report LTM in ceramic samples with GS less than 1000 nm. For samples with GS more than 1200 nm, only a single sharp maximum in the $\rho(T)$ curve has been reported. The SEM pictures of the $La_{0.67}Ca_{0.33}Mn_{1-x}Ru_xO_3$ ($x \leq 0.10$) compounds are shown in **FIG.9** It is seen that grain size (GS) of the undoped (x = 0) compound is ~ 20,000 nm. Though the GS marginally decreases with Ru substitution, it is typically in the range of 16000-18000 nm. An overall improvement in the grain connectivity is observed for the substituted systems. Other studies show the LTM to shift to lower temperatures with the decrease of GS. For instance, Zhang et al.[55] have reported a shift of ~ 25 K as the GS decreased from 1000 to 50 nm and the LTM occurs at ~ 250 K for sample with GS of 155 nm. In the case of Ru substituted compounds, LTM shifts at a rate of ~ 16 K/at.% and it is believed that such a large shift cannot arise from the decrease in GS. For example, the LTM for x = 0.10 occurs at ~ 120 K. If this is to arise from the GS effect, the GS should be unphysically small.[55] The residual resistivity is shown to increase by a few orders of magnitude with the decrease of GS below 1000 nm. However, the observed increase in $\rho_o$ in our Ru substituted compounds is far less than this. Another important point to be noted is that MR(T) curves under 5 T shows a sharp maximum close to $T_{MI1}$ and a broad maximum close to $T_{MI2}$ (Error! Reference source not found.). This is in contrast to the reports wherein the latter peak is absent in the MR curve for compounds exhibiting double maxima in R(T, H) curve due to GS.[42,56] Additionally, the magnitude of MR at its peak is reported to be sensitive to GS and falls sharply with the decrease of GS. Compounds with smaller GS are shown

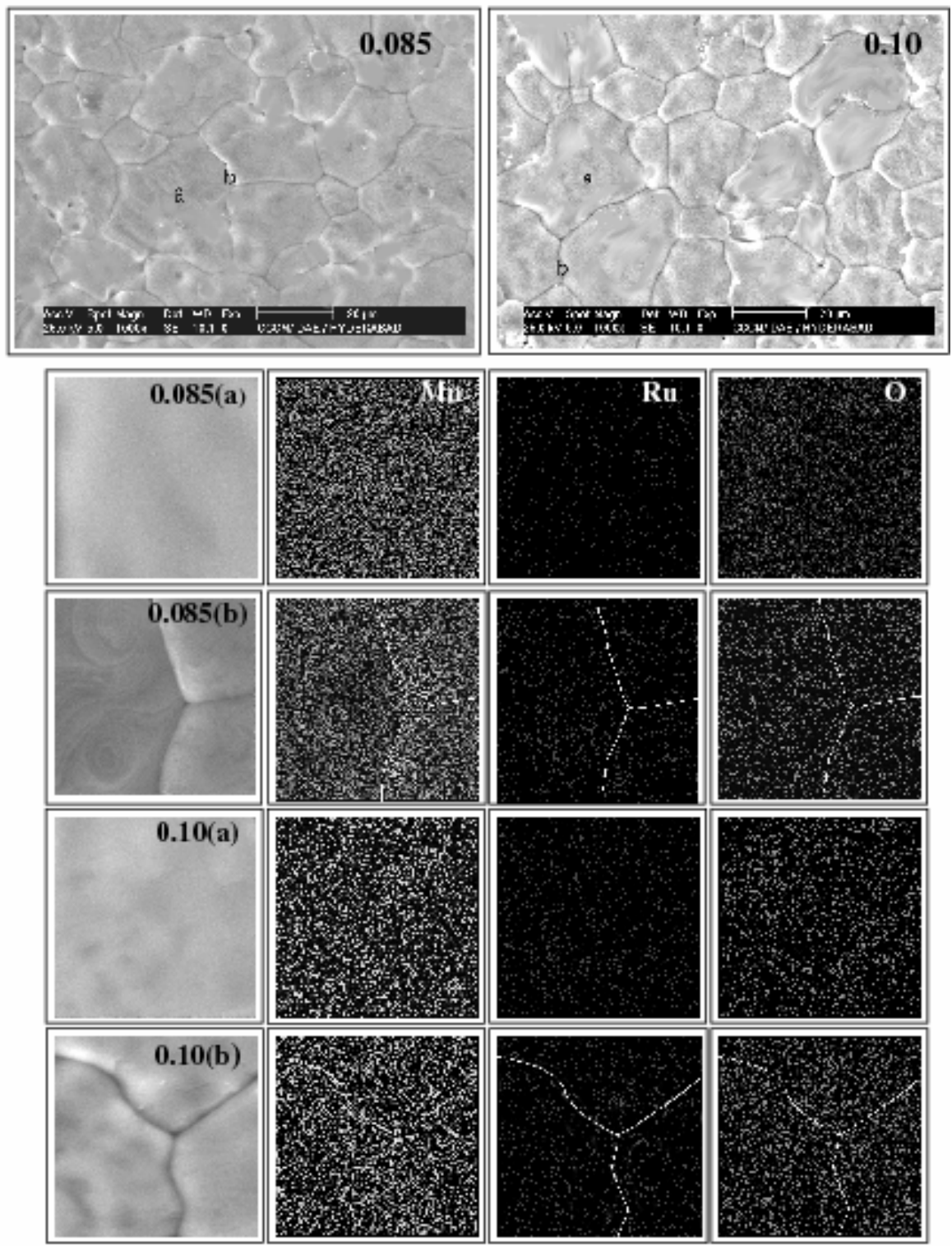

**FIG.10**: EDAX map of Mn, Ru and O in the selected regions (a) within the grain and (b) across the grains of $La_{0.67}Ca_{0.33}Mn_{1-x}Ru_xO_3$ (x = 0.085 and 0.10) compounds. For a comparison, the corresponding SEM micrographs are also shown in the topmost panel

to exhibit a MR peak value less than 20 %,[45] whereas in x = 0.10, the highest Ru concentration of present study, MR peak values as high as 66 % and 45 % are observed at $T_{MI1}$ and $T_{MI2}$ respectively (**FIG.8**). The foregoing discussion clearly indicates that GS is certainly not the cause/origin for the observed LTM in the $\rho$ (T) curves of the Ru substituted $La_{0.67}Ca_{0.33}MnO_3$.

Presence of double maxima feature in the $\rho(T)$ has been reported for the Ln-site substituted LnMnO$_3$ system with $Ce^{4+}$ ion.[43, 44] However, it has been shown that compounds were of multi-phase nature evidenced by the presence of additional peak(s) in XRD pattern.[44] From the high statistics XRD pattern collected for Ru substituted compounds, the presence of impurity phase with volume fraction more than 1 % is excluded and no peak was left un-indexed (**FIG.1**). The lattice parameters and the unit cell volume are found to increase linearly over the entire range of Ru compositions (**FIG.2**). Additionally, the transition temperatures determined from the resistivity, ac susceptibility (**FIG.5**), DSC measurements (**inset of FIG. 6**) are in close agreement and decrease linearly over the entire composition range ruling out impurity phase formation. In order to see the chemical homogeneity of the samples, EDAX mapping of Mn, Ru and O is carried out at randomly selected regions within the grain and across the grains. As a representative of the series, elemental distribution of $La_{0.67}Ca_{0.33}Mn_{1-x}Ru_xO_3$ (x = 0.085 and 0.10) compounds are shown in **FIG.9** The mapping shows a uniform distribution of Mn, Ru and O ions within as well as across the grains. This indicates that the compounds are chemically homogeneous within the limitation of the EDAX method. Thus, we rule out chemical inhomogeneity as being the cause of LTM in the $\rho(T)$.

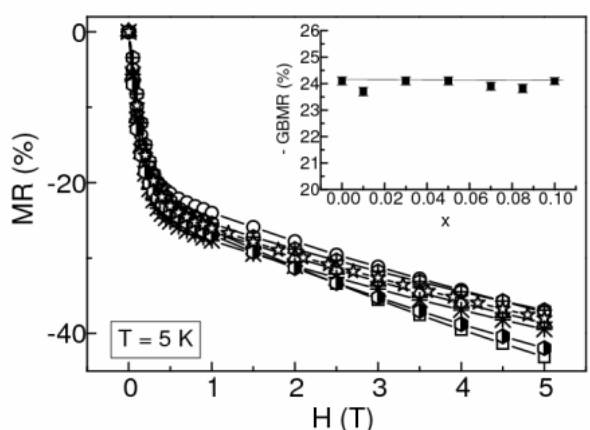

**FIG. 11:** Field dependence of MR (MR(H)) at 5 K of $La_{0.67}Ca_{0.33}Mn_{1-x}Ru_xO_3$ compounds (x = 0 (□), 0.01 (○), 0.03 (△), 0.05 (✳), 0.07 (⊕), 0.085 (☆) and 0.10 (◐)). Inset shows the variation of GBMR (in %) at 5 K as a function of x of the compounds.

As shown in other works,[47, 48, 57] GB can also lead to a broad LTM. The crucial measurement to estimate the GB effects is the evolution of MR under moderate magnetic fields (MR(H)) at different temperatures. MR Vs H curve at 5 K (**FIG. 11**) exhibits two slopes- a sharp linear fall for field less than 1 T followed by much slower decrease above 1 T. The GB contribution to MR (GBMR) was estimated by extrapolating high field MR and finding its intercept at zero field. GBMR for all the compounds at 5 K is found to be ~ 24 % with systematic variation of Ru concentration. (**inset of FIG.11**). The insensitiveness of GBMR to the Ru composition indicates that GB is not much altered upon Ru substitution to warrant a LTM.

The effect of oxygen off-stoichiometry ($\delta$) in $La_{0.7}Ca_{0.3}MnO_{3-\delta}$ is reported to be as dramatic as any other extrinsic factor mentioned above.[49] Apart from shifting MIT to lower temperatures, it is also reported to result in two maxima in the $\rho(T)$ curves. Width of the transition is reported to be $\delta$ dependent, increasing with reducing oxygen content. Larger values of $\delta$ are shown to result in structural transition and render the compound into a semiconductor. Trukhanov. et. al[58] also have shown that for $\delta > 0.06$, ferromagnetic-metallic ground state of $La_{0.7}Ca_{0.3}MnO_{3-\delta}$ transforms to a cluster-glass insulator. The TGA curves of $La_{0.67}Ca_{0.33}Mn_{1-x}Ru_xO_3$ ($0 \leq x \leq 0.10$) compounds are shown in **FIG.12.** The oxygen stoichiometry was found to show a marginal variation from the nominal composition of 3. However, no systematics could be deduced with the Ru composition. If $\delta$ is to be the cause of LTM of the Ru substituted compounds, the expected $\delta$ value for x = 0.10 is 0.19. On the other hand, the estimated $\delta$ for the x = 0.10 compound is ~ 0.01. Going by the reported works, it is expected that x $\geq$ 0.05 to undergo a structural transition and to exhibit semiconducting behavior down to 5 K. However, Ru substituted compounds are found not to alter orthorhombic structure (Pnma symmetry) over the entire range of substitution and retain metallic ground state. Also, such large value of $\delta$ should have resulted in drastic increase in the value of $\rho_o$, typically by few orders of magnitude. No such large change in $\rho_o$ is observed in $La_{0.67}Ca_{0.33}Mn_{1-x}Ru_xO_3$ (0 $\leq$ x $\leq$ 0.10) compounds.

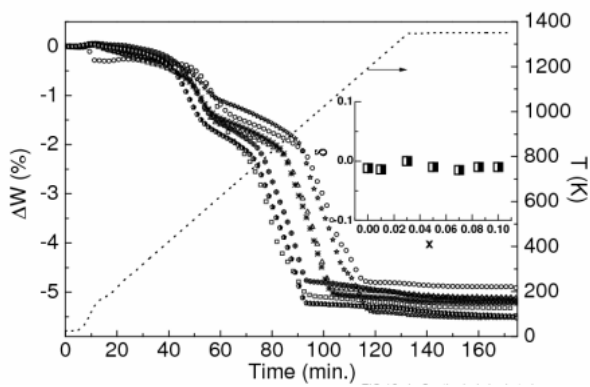

**FIG.12:** TGA curves of $La_{0.67}Ca_{0.33}Mn_{1-x}Ru_xO_3$ compounds for x = 0 (□), 0.01 (○), 0.03 (△), 0.05 (✳), 0.07 (⊕), 0.085 (☆) and 0.10 (◐). Inset shows the oxygen off-stoichiometry ($\delta$) as a function of x of the compounds.

From the preceding discussions, it is established that the LTM in the $\rho$(T) is not due to any of the extrinsic factors. The most striking feature is the presence of double maximum in the MR(T) curve about their respective metal-insulator transition temperatures. Though the low temperature MR peak is rather weak for lower x, it becomes dominant and distinct for x = 0.10. The LTM in the MR(T) not only excludes the extrinsic factors to be the cause of the LTM in the $\rho$(T), but also establishes the fact that it is intrinsic for the $La_{0.67}Ca_{0.33}Mn_{1-x}Ru_xO_3$ system. Having established so, it was proposed for the first time that Ru substituted compounds undergo a magnetic phase separation, different from electronic phase separation occurring in other systems where, AFM-I and FM-M coexist at low temperatures.[6,59] The volume fraction of these two phases in the electronic phase separated system could easily be altered by a suitable thermodynamic variable such as magnetic field, pressure or by chemical doping as in the case of $(La_{1-x}Pr_x)_{0.67}Ca_{0.33}MnO_3$ system.[6] With the increase of the AFM-I phase, ie by increasing the Pr concentration, $\rho_o$ is found to increase by several orders of magnitude. In contrast, in the case of Ru substituted $La_{0.67}Ca_{0.33}MnO_3$ compounds, $\rho_o$ increases by a factor of 5.[32] Also in the case of electronic phase separation, only a single maximum in the resistivity curves is expected, since out of the two phases, the FM phase alone is metallic in nature. Thus it is concluded that the phase separation encountered in Ru substituted $La_{0.67}Ca_{0.33}MnO_3$ compounds is not an electronic phase separation, since both the phases are FM-M in their ground state. Nevertheless, they could still differ in their magnetic properties. The difference in the electrical transport and magnetic properties, influenced by the strength of DE interaction can be rationalized in the following way.

Drawing analogy with the $Fe^{3+}$ substituted $La_{0.67}Ca_{0.33}MnO_3$ compounds having a $dT_c/dx$ of ~18 K/at.%,[26] and also in light of XPS studies on Ru doped La-Sr-Mn-O system by Krishnan et.al,[38] and XMCVD studies by Weigand et. al,[60] mixed valence state of Ru, viz, $Ru^{4+}$ (iso-electronic to $Mn^{3+}$) and $Ru^{3+}$ (iso-electronic to $Fe^{3+}$) has been inferred. The double maxima in the $\rho$(T) curve corresponds to the evolution of two ferromagnetic-metallic phases: $Ru^{4+}$ rich regions with local ferromagnetic coupling of Ru with the neighboring Mn ions and $Ru^{3+}$ rich regions with local antiferromagnetic coupling of Ru with neighboring Mn ions. Of the two FM phases, the latter is a weaker ferromagnetic phase. The local ferromagnetic coupling between $Ru^{4+}$ and neighboring Mn spins favors the Double Exchange (DE) interaction strength leading to a small $dT/dx$ (~ 3 K/at.%) associated with $T_{MI1}$, $T_c$ and $T(C_p)$. On the other hand, $Ru^{3+}$ ion with its half filled electronic configuration cannot participate in DE interaction. Moreover, its local AFM coupling with neighboring Mn spins in the presence of competing antiferromagnetic superexchange interactions weakens the DE interaction strength which in turn results in a larger suppression rate ($dT_{MI2}/dx$) of ~ 16 K/at.% comparable to that of $Fe^{3+}$ substituted CMR manganites. Progressive lowering of $T_{MI1}$ and $T_{MI2}$ with Ru concentration indicates that the corresponding phases are getting progressively enriched in $Ru^{4+}$ and $Ru^{3+}$, respectively. The enrichment of $Ru^{3+}$ and/or an increase in the volume fraction of the weaker ferromagnetic (poor conducting) phase is expected to result in the observed rise in the value of $\rho_0$ of the substituted compounds. The origin of the double maxima in $\rho$(T) can then be understood in the context of magnetic phase separation. More detailed discussion of the magnetic phase separation scenario is published elsewhere.[32,33] As the sample is cooled down from higher temperatures (i.e. paramagnetic insulating phase) a ferromagnetic phase (FM-M1) evolves within a paramagnetic (PM) matrix, having appreciable resistivity contrast between the phases. Upon reaching a critical volume fraction $V_C$, FM-M1 establishes a percolative path and the resistivity drops drastically resulting in a maximum in $\rho$(T). On further cooling, the PM phase itself undergoes a para to ferromagnetic transition. As the system is percolatively conducting below the $T_{MI1}$, magnetic transition of the PM phase is not expected to produce drastic drop in the resistivity, as seen for $T_{MI1}$. The magnetic field results in an expected shift of $T_{MI1}$ to higher temperature. Such a shift is not expected in the case of $T_{MI2}$ for the reason that the magnetic transition of PM phase takes place within a conducting FM-M1 matrix with a low conductivity contrast. This magnetoresistance signal corresponding to the LTM in zero field resistivity grows into a broad maximum under a magnetic field of 5 T but far smaller compared to that of $T_{MI1}$. These features, negligible shift of $T_{MI2}$ and substantially reduced MR maximum associated with $T_{MI2}$ support the conjecture of a magnetic phase separation occurring in $La_{0.67}Ca_{0.33}Mn_{1-x}Ru_xO_3$ compounds.

## VI. SUMMARY

Possible role of the extrinsic factors such as synthesis conditions, chemical inhomogeneity, grain size, grain boundary effects and oxygen-off stoichiometry for the second low temperature maximum on the electrical transport of Ru substituted $La_{0.67}Ca_{0.33}MnO_3$ are analysed in light of the SEM, EDAX, field dependent magnetoresistance and thermogravimetric studies. In our opinion, synthesis conditions do not play a role on the electrical transport of the Ru substituted system. Typical grain size of ~ 18,000- 20,000 nm as estimated from the SEM micrographs are not so small as to warrant a LTM. Absence of any additional peaks in high statistics XRD patterns, linear systematic increase of the unit cell parameters, close matching of the transition temperatures in the resistivity, ac susceptibility and specific heat, the linear systematic decrease in the transition temperatures with Ru composition and homogeneous distribution of Mn, Ru and O at arbitrarily selected areas within the grain and across the grains rule out the chemical inhomogeneity in the samples. The insensitiveness of grain boundary contribution of MR at 5 K to the Ru composition indicate that grain boundary is not altered much on Ru substitution to warrant a LTM in the electrical transport. The oxygen stoichiometry of the compounds are close to the nominal composition 3. These results not only excludes the extrinsic factors, but also unambiguously establish that the double metal-insulator transitions, both exhibiting magnetoresistance is intrinsic to Ru substitution. The origin of the double maximum is understood in terms of the magnetic-phase separation scenario involving the co-existence of two ferromagnetic-metallic phases in its ground state.

**ACKNOWLEDGEMENT**

We wish to thank Dr. T. Geethakumary for providing the ac susceptibility set up and Mr. P. K. Ajikumar for the thermogravimetric studies. One of the authors, LSL also acknowledges Council of Scientific and Industrial Research, India, for an award of a Senior Research Fellowship.